\documentclass [twocolumn, aps, showpacs] {revtex4}
\usepackage{graphicx}
\begin{document}
\draft
\title {Dissipation peak as an indicator of sample inhomogeneity in solid $^4$He oscillator experiments}
\author {David A. Huse and Zuhair U. Khandker}
\address{Department of Physics, Princeton
University, Princeton, NJ 08544}
\begin{abstract}
A simple phenomenological model is developed for the recent
torsional oscillator experiments on solid $^4$He.  Within this
model, for a homogeneous sample there is a specific quantitative
relation between the change in the oscillator's frequency and its
maximum damping at the apparent supersolid transition.  Much of the
published data do not satisfy this relation, indicating that the
dissipation peaks in those samples are strongly inhomogeneously
broadened.
\end{abstract}
 \maketitle

Kim and Chan \cite{kimchan1, kimchan2, kimchan3}, and more recently
Rittner and Reppy \cite{randr, randr2}, report apparent
superfluidity in torsional oscillator experiments on {\it solid}
$^4$He.  They freeze $^4$He within a high-Q oscillator and measure
the period and the damping of the oscillator as a function of the
temperature $T$. As $T$ is decreased, in a temperature range near
50-200 mK the oscillator's period is observed to decrease,
indicating a reduction of the effective moment of inertia of the
oscillator at low $T$. This is interpreted as the development of
superfluidity within the solid, with the superfluid component
remaining stationary and thus no longer contributing to the moment
of inertia of the oscillator at low $T$.  In the temperature range
where the oscillator's period is most strongly
temperature-dependent, the oscillator's damping has a maximum
(equivalently, a minimum in the quality factor $Q$ of the
oscillator's resonance).

A simple phenomenological model of these experiments is of a
two-component compound torsional oscillator:
$$(I-I_s)\frac{d^2\theta}{dt^2}=-G\theta-\Gamma(T)I_s\frac{d(\theta-\phi)}{dt}$$
$$I_s\frac{d^2\phi}{dt^2}=-\Gamma(T)I_s\frac{d(\phi-\theta)}{dt}~.$$
The {\it relative} angular motion of the two components,
$d(\phi-\theta)/dt$, is damped at a rate $\Gamma(T)$ that is
strongly $T$-dependent.  This relative motion experiences no
restoring force. This motion represents the superfluid component
($\phi$) moving with respect to the ``normal'' solid ($\theta$),
with the damping likely coming from the dissipative motion of
thermally-excited vortices in the superfluid. At high $T$ the
damping is strong so $\Gamma(T)\gg\omega\approx\sqrt{G/I}$, where
$\omega$ is the resonant frequency.  The two components are then
effectively locked together with total moment of inertia $I$. At low
$T$ the damping is weak or zero, $\Gamma(T)\ll\omega$, so the
smaller, nominally superfluid component, with moment of inertia
$I_s\ll I$ and angle $\phi$, remains essentially at rest (the
$T$-dependence of $I_s$ is assumed to be much weaker than that of
$\Gamma$, and thus is ignored). The resulting fractional change in
the resonant frequency from high $T$ to low $T$ is
$\frac{\Delta\omega}{\omega}\approx\frac{I_s}{2I}\ll 1$, due to the
decrease of the moving moment of inertia from $I$ to $I-I_s$. As
$\Gamma(T)$ passes through $\omega$, this simple model has a
dissipation peak, with the superfluid component then moving
substantially both in the lab frame and relative to the solid. Let
the magnitude of this damping peak be measured by $\Delta(Q^{-1})$,
the increase in the oscillator's inverse quality factor $Q^{-1}$ at
this peak. Our simple model gives
$\Delta(Q^{-1})\approx\frac{I_s}{2I}$.

This simple result,
$\Delta(Q^{-1})\cong\frac{\Delta\omega}{\omega}$, is roughly
consistent with torsional oscillator experiments on superfluid {\it
liquid} $^4$He films by Bishop and Reppy \cite{breppy}.  This result
was also obtained from a more realistic model of superfluid liquid
films within a Kosterlitz-Thouless theory by Ambegaokar {\it et al.}
\cite{ahns}.

However, the ratio of the magnitude of the dissipation peak
$\Delta(Q^{-1})$ to the fractional frequency shift
$\frac{\Delta\omega}{\omega}$ varies widely among the published data
on {\it solid} $^4$He oscillator experiments \cite{kimchan2,
kimchan3, randr, randr2}. This is likely due to inhomogeneous
broadening of the dissipation peak, from local ``decoupling''
transitions occurring at different temperatures in different parts
of a macroscopically inhomogeneous polycrystalline solid, with the
superfluidity arising on a network of grain boundaries or other
defects in the solid \cite{ns,pea}. The total frequency shift
$\Delta\omega$ from high to low $T$ is set by the total amount of
moment of inertia that decouples from the oscillator. For strong
inhomogeneity, the $T$ range over which this happens reflects the
variation of the local transition temperatures. The damping at
temperature $T$ is only due to those parts of the solid that are
passing through their transition at that $T$ so that they have a
local $\Gamma(T)$ that is near $\omega$. Thus the magnitude of the
dissipation peak $\Delta(Q^{-1})$ in an inhomogeneous solid is
reduced from $\frac{\Delta\omega}{\omega}$ by the fraction of the
sample that is passing through its local decoupling transition and
thus contributing to the damping at the dissipation peak.

By this criterion, the data of Kim and Chan \cite{kimchan2,
kimchan3} suggest they had much more inhomogeneous samples than
those of the first paper of Rittner and Reppy \cite{randr}. Among
those published, the result of Kim and Chan that appears the most
homogeneous is that for pressure 30 bar shown in Fig. 2 of Ref.
\cite{kimchan3} (we restrict our attention to the lowest amplitude
of oscillation). This run shows $\Delta(Q^{-1})\cong 3\times
10^{-6}$, about a factor of 6 or 7 smaller than its
$\frac{\Delta\omega}{\omega}\cong 20\times 10^{-6}$.  In other
examples from Kim and Chan \cite{kimchan2, kimchan3} this factor
ranges up to about 100. In their first paper, Rittner and Reppy
\cite{randr}, on the other hand, show data closer to the homogeneous
expectation of $\Delta(Q^{-1})\cong\frac{\Delta\omega}{\omega}$.
E.g., their Fig. 2 shows $\Delta(Q^{-1})\cong 12\times 10^{-6}$ and
$\frac{\Delta\omega}{\omega}\cong 18\times 10^{-6}$.  (The sample
they label ``First Run'' even has $\Delta(Q^{-1})$ roughly 3 times
{\it larger} than $\frac{\Delta\omega}{\omega}$, which is peculiar.)
In Rittner and Reppy's more recent paper \cite{randr2}, they
intentionally made quite disordered samples, and these samples do
show much weaker dissipation peaks indicating they are also quite
inhomogeneous, as might be expected.

Another important difference between the published results from
these two labs is that for Rittner and Reppy \cite{randr} the
transition disappeared in one crystal after annealing, while Kim and
Chan \cite{kimchan3} saw no large changes due to any annealing of
their samples.  This difference might also be related to differences
in the sample inhomogeneity.  One possible cause of sample
inhomogeneities that would survive annealing is stresses due to
temperature gradients that are present in the sample cell during the
annealing and/or during the post-anneal cooling.


We thank Phil Anderson, Bill Brinkman, Moses Chan, Tony Clark, and
John Reppy for discussions.  Support was from the NSF through MRSEC
grant DMR-0213706.

\end{document}